\documentclass[12pt,english]{article}
\usepackage[T1]{fontenc}
\usepackage[latin9]{inputenc}
\usepackage{setspace}
\makeatletter
\usepackage{aStyle}

\makeatother

\usepackage{babel}
\begin{document}
\begin{doublespace}
\begin{center}
\textbf{\Large{}The Epistemology of Contemporary Physics:\\Classical
Mechanics II}\vspace{-1.3cm}
\par\end{center}
\end{doublespace}

\begin{center}
Taha Sochi (Contact: ResearchGate)\vspace{-0.4cm}
\par\end{center}

\begin{center}
London, United Kingdom
\par\end{center}

\noindent \textbf{Abstract}: In this paper of ``The Epistemology
of Contemporary Physics'' series we investigate Newton's third law
and discuss and analyze its epistemological significance from some
aspects with special attention to its relation to the principle of
conservation of linear and angular momentum. The main issue in this
investigation is the potential violations of this law according to
the claims made in the literature of mainstream physics. This issue
may cast a shadow on the validity of classical mechanics, and its
Newtonian formulation in particular, formally and epistemologically
and could have important implications and consequences on contemporary
physics in general. However, what is more important about this issue
from our perspective is the lack of clarity, comprehensibility and
coherence in the investigation and analysis of this issue and its
implications marked by the absence of appropriate conceptual and epistemological
frameworks to deal with this issue properly and systematically. As
a result, what we find in the literature is a collection of contradicting
views which are mostly based on personal choices and preferences and
selective or biased theoretical analysis with the lack of proper experimental
verification and substantiation.\vspace{0.3cm}


\clearpage{}

\tableofcontents{}

\clearpage{}

\section{Introduction}

Newton's third law is one of the pillars of classical mechanics in
its Newtonian formulation (and possibly further). It may also be seen
as the most important and physically significant law among Newton's
laws of motion (as discussed already in \cite{SochiEpistClass12024})
due mainly to its link to the principle of conservation of linear
and angular momentum (see for instance \cite{TaylorBook2005}) which
is one of the main pillars of all physics. In fact, some scholars
even consider this law as one of the most important laws of Nature
(see for instance \cite{Cornille1999}).

Nonetheless, this law is one of the most troubling parts and problematic
aspects of the Newtonian formulation of classical mechanics (and possibly
classical mechanics in general and even beyond). This is not only
because of its rather epistemological obscurity (see \cite{SochiEpistClass12024})
but also because of its (claimed) formal violations and exceptions
which cast a shadow on its validity and hence on its status as a real
physical law (as well as putting question marks on its epistemological
significance and its relation to the conservation of momentum among
other implications and consequences). However, it is important to
note that the blame in these claimed violations and exceptions may
be put (in some cases at least) on other physical theories and aspects
(like Lorenz electrodynamics theory) rather than on Newton's third
law itself.\footnote{This issue should be investigated further in due course within this
series.} Moreover, such violations are denied altogether by some scholars
who justify their denial by certain conceptualizations or formulations
(or even novel theories within or outside the mainstream physics).

The problematic nature of Newton's third law (or perhaps the problematic
aspects surrounding this law which possibly relate to other physical
theories and aspects as indicated already) was observed rather early
in the history of contemporary physics. For example, Henri Poincare
noticed that (see pages 194-195 of \cite{PoincareBook1905}):

The most satisfactory theory is that of Lorentz; it is unquestionably
the theory that best explains the known facts, the one that throws
into relief the greatest number of known relations, the one in which
we find most traces of definitive construction. That it still possesses
a serious fault I have shown above. It is in contradiction with Newton's
law that action and reaction are equal and opposite - or rather, this
principle according to Lorentz cannot be applicable to matter alone;
if it be true, it must take into account the action of the ether on
matter, and the reaction of the matter on the ether. Now, in the new
order, it is very likely that things do not happen in this way. (End
of quote)

In more recent times a number of problematic (or potentially problematic)
aspects related to Newton's third law are similarly identified (or
indicated or investigated) by a number of scholars. Some of the common
views in the literature about this issue are (noting that some of
these views may originate from the same cause):

\noindent $\bullet$ Newton's third law is violated or has no place
in relativistic mechanics (see for instance \cite{FrenchBook1968,TaylorBook2005}).

\noindent $\bullet$ Newton's third law is violated in electrodynamics
(see for instance \cite{GriffithsBook2013}).

\noindent $\bullet$ Newton's third law is violated by Lorentz force
law (see for instance \cite{Cornille1995,Cornille1999}).

\noindent $\bullet$ Biot-Savart law does not obey Newton's third
law (see for instance \cite{Mathur1941,Christodoulides1988}).

\noindent $\bullet$ Newton's third law is violated in the interaction
(or rather relationship) between the (Newtonian) space and the material
objects (see for instance \cite{JammerBook1993}).

\noindent $\bullet$ Inertial forces do not obey Newton's third law
(see for instance \cite{Irodov2002}).

\noindent $\bullet$ Non-equilibrium forces (in certain physical systems)
violate Newton's third law (see for instance \cite{DzubiellaLL2003}).

Our intention in the present paper is to investigate in rather sufficient
details the significance and limitations of Newton's third law and
its (potential) violations where we focus in this respect on analyzing
its relationship to the principle of conservation of linear and angular
momentum. The main purpose of all this is to asses the effect of the
(potential) violations of Newton's third law on the validity and applicability
of the Newtonian formulation of classical mechanics, and hence the
effect of this on the classical mechanics and contemporary physics
in general (with special attention and consideration to the epistemological
and interpretative aspects of these issues).

However, our investigation should also reveal and highlight the messy
situation of the investigations of these issues and the lack of consistent
and objective scientific methodology in most of the approaches that
deal with these issues in the literature. The situation is generally
surrounded with ambiguities, misconceptions, randomness, selectivity
and lack of sufficient experimental substantiation and verification.

The structure of this paper is that we start with a preliminary section
(where we investigate some introductory issues related to our main
subject). We then investigate in the subsequent section some instances
of violation (or rather claimed or tentative or potential violation)
of Newton's third law which we find in the literature of mainstream
physics (noting that there are other instances of claimed violations
in the literature of the so-called ``fringe science'' which we do
not consider in this paper). We then discuss briefly (in another section)
some general remarks related to our main investigation before we outline
in a ``Conclusions'' section the main results and conclusions that
we obtained from our investigation in the present paper.

\section{Preliminaries}

We investigate in this preliminary section some important issues related
to our main subject (as explained in the Introduction).

\subsection{Relationship between Newton's Third Law and Conservation of Momentum}

According to the literature of mainstream physics, the physical essence
of Newton's third law is the conservation of momentum and this is
shown, for instance, through the derivation of the principle of conservation
of momentum from Newton's third law (which can be found in elementary
physics textbooks and even in some secondary school curricula).\footnote{In fact, the conservation of both linear and angular momentum can
be derived from Newton's third law noting that this law consists of
three main facts about the action-reaction forces on the two interacting
objects: (\textbf{a}) the equality of the two forces in magnitude,
(\textbf{b}) being in opposite directions, and (\textbf{c}) being
along the line joining the two objects. Further clarifications about
this issue will follow.} 

However, despite this intimate relationship between Newton's third
law and the conservation of momentum they are not equivalent,\footnote{Statements that assert the equivalence of Newton's third law and the
conservation of momentum can be found in the literature. For example,
we read in Taylor (see page 21 of \cite{TaylorBook2005}) ``conservation
of momentum and Newton's third law are equivalent to one another''.
However, the meaning of ``equivalent'' is rather different to what
we mean here.} neither formally nor epistemologically (and this seems obvious from
analyzing their contents and implications). For example, the derivation
of the conservation of (linear and angular) momentum from Newton's
third law usually depends on the use of Newton's second law (in its
linear and angular forms) which is not incorporated in the conservation
of momentum or part of it (see for instance \cite{FrenchBook1971,TaylorBook2005}).\footnote{In fact, the non-equivalence should be aggravated if we adopt the
view that Newton's second law is a definition in some sense (see \cite{SochiEpistClass12024}).} On the other hand, the conservation of momentum is more general in
theory and application than Newton's third law (at least among the
majority of mainstream physicists) since the conservation of momentum
is valid and applicable in theories and branches other than classical
mechanics and its Newtonian formulation and beyond their domain of
validity (in fact there seems to be a general consensus among physicists
that the conservation of momentum is valid and applicable in all physics
and hence it is one of the most fundamental laws or principles of
Nature). This can be concluded, for instance, from the fact that the
conservation of momentum in modern physics is usually obtained from
theoretical arguments based on the properties of space\footnote{\label{fnConMomSym}The conservation of linear momentum is supposed
to be based on the homogeneity of space while the conservation of
angular momentum is supposed to be based on the isotropy of space.} and related symmetries (noting that these arguments are more general
than classical mechanics and hence they extend beyond its domain of
validity; moreover they are not based on Newton's third law).

Nevertheless, if the derivation of the conservation of momentum (within
classical mechanics or rather its Newtonian formulation) is based
exclusively on Newton's third law then we can say that the conservation
of momentum is violated within the framework of Newtonian formulation
of classical mechanics, and hence if the conservation of momentum
is a fundamental principle of all physics that cannot be violated
(as suggested already) then this means that classical mechanics (in
its Newtonian formulation at least) is incorrect from this aspect
or at least it requires rectification and correction or amendment
(such as by imposing certain extra conditions on its validity and
applicability). On the other hand, if violation to the conservation
of momentum is allowed (which seems to be a bizarre view) then we
may be able to rectify this embarrassing situation in classical mechanics
by putting some restrictions and conditions on Newton's third law
(noting that our loss by accepting a violation to the conservation
of momentum should be greater than any supposed gain by this rectification).

\subsection{Importance of Newton's Third Law}

The importance of Newton's third law was investigated in a previous
paper of this series (see \cite{SochiEpistClass12024}) and also implied
by the discussion of the previous subsection. What we need to add
here is that the importance of Newton's third law should be demonstrated
and reflected in two main aspects:
\begin{enumerate}
\item The importance of this law in itself within the framework of classical
mechanics (and its Newtonian formulation in particular).
\item The importance of this law in physics in general mainly through its
relation to the principle of conservation of (linear and angular)
momentum (and possibly through other implications and consequences).
\end{enumerate}
\vspace{0.1cm}So, from the first aspect any potential violation to
this law should affect classical mechanics but not necessarily other
parts and disciplines of physics (and the conservation of momentum
in particular), while from the second aspect any potential violation
could have far reaching consequences on the entire physics. The second
aspect should depend primarily on the relationship between Newton's
third law and other laws and principles of physics (particularly the
principle of conservation of momentum), where this relationship is
essentially determined by the presumed position and role of Newton's
third law within the contemporary physics in general. In fact, there
are conflicting views and opinions in this regard where some scholars
grant this law a central position and role within the contemporary
physics in general (due mostly to its supposedly strong relationship
with the principle of conservation of momentum in general) while other
scholars restrict its role (or main role) to classical mechanics (and
the Newtonian formulation in particular).

Anyway, the first aspect may not represent a serious problem due to
the already-imposed limitations on the validity and applicability
of classical mechanics (which we discussed in \cite{SochiEpistClass12024})
and considering that classical mechanics is generally seen as an approximate
theory or a limit to other fundamental theories, and hence any violation
will not introduce a fundamental change on the status of classical
mechanics (and its Newtonian formulation in particular) from this
perspective. This is unlike the second aspect since it extends beyond
classical mechanics to reach (at least potentially) other physical
theories and disciplines and hence it could affect physics fundamentally
and in general (noting in particular its potential effect on the conservation
of momentum which is supposedly a fundamental principle of all physics).

We should also mention (with regard to the first aspect) that any
violation of Newton's third law should affect the Newtonian formulation
of classical mechanics but should not affect the other formulations
of classical mechanics directly since Newton's third law is proprietary
to the Newtonian formulation. However, due to the supposed equivalence
of these formulations to the Newtonian formulation (as explained in
\cite{SochiEpistClass12024}) these formulations could be affected
indirectly through their implications and consequences which are related
to this equivalence. Accordingly, violations to Newton's third law
should affect (in this sense and capacity) classical mechanics in
general.

\subsection{Weak and Strong Forms of Newton's Third Law}

According to the literature of physics, Newton's third law has two
forms: weak and strong. In fact, Newton's third law consists of three
main components or ingredients related to the attributes of the action
and reaction forces which act on the two interacting objects and their
relationship:

\noindent $\bullet$ These forces are equal in magnitude.

\noindent $\bullet$ These forces are opposite in direction.

\noindent $\bullet$ These forces are acting along the line joining
the two interacting objects (or rather particles).

\noindent Accordingly, while the weak form of Newton's third law consists
of the first two components only, the strong form consists of all
these three components. 

The following points are worth noting in this regard:
\begin{enumerate}
\item The violation of Newton's third law may occur (in theory and hypothetically)
by violating both forms of Newton's third law (i.e. by violating at
least one of the first two components and possibly the third component
as well) or by violating the strong form only (i.e. by violating the
third component only). In fact, the literature seems to contain claims
of potential violations of both these forms in various physical situations
and circumstances or scenarios where the violation of each form occurs
independently, i.e. there are examples of potential violation of the
strong form only and other examples of potential violation of both
forms (noting that violation of the weak form implies, in some sense,
violation of the strong form).
\item We should notice that while some scholars consider both forms as being
valid and legitimate (in the sense that they represent a sort of ``independent''
physical laws, or rather independent representations of Newton's third
law, that apply in different situations and circumstances), other
scholars seem to consider only one of these forms as being valid and
legitimate (and hence if the weak form is the actual and real representation
of Newton's third law then there is no violation to Newton's third
law by violating the third component,\footnote{We note that violating the third component in this case may imply
violation of some physical law or principle (other than Newton's third
law), but this should be another issue not related to Newton's third
law according to this representation of Newton's third law.} while if the strong form is the actual and real representation of
Newton's third law then violating any one of the three components
is a violation to Newton's third law with no distinction between which
component is the violated one).\\
Anyway, it seems that most scholars (see for instance $\S$ 1.5 of
\cite{TaylorBook2005}) adopt the view that Newton's third law is
actually the weak form, and hence the strong form (or rather its content)
is restricted to central forces. This may seem logical because the
conservation of angular momentum (which is based on the third component)
is restricted to central forces (see for instance pages 94-95 of \cite{TaylorBook2005}).
This should also be inline with the common position in the literature
about the nature of the relationship between the weak/strong form
of Newton's third law and the conservation of linear/angular momentum
(as indicated earlier), i.e. the weak form implies the conservation
of linear momentum while the strong form implies the conservation
of angular momentum.
\item Noting that the third component is usually used in the derivation
of the conservation of angular momentum, the violation of the strong
form (of Newton's third law) specifically and exclusively may be associated
with the violation of the conservation of angular momentum. If so
then violating the weak form implies violation of conservation of
linear momentum, while violating the strong form (specifically) implies
violation of conservation of angular momentum.\footnote{When we say ``implies'' it should mean ``potentially implies''
noting that we are not aware of physicists who accept such implications.}
\end{enumerate}

\subsection{\label{subsecTypicalExample}Typical Example of Claimed Violation
of Newton's Third Law}

To clarify the situation further let have a simple example that demonstrates
a typical and common instance of claimed violation of Newton's third
law\footnote{In fact, this typical and common example of violation is related to
electrodynamics and relativistic mechanics (see $\S$ \ref{subsecElectrodynamics}
and $\S$ \ref{subsecRelativistic}) which are the main fields in
which claimed violations of Newton's third law are common.} (noting that there are many other examples in the literature about
claimed violation of Newton's third law which are related to various
subjects and fields as will be investigated in detail later on; see
$\S$ \ref{secInstancesViolation}).\footnote{In fact, we are using simple (and rather non-technical) language in
our explanation to this simple example to demonstrate some essential
points and aspects that we need to examine in our subsequent discussions.
It should be noted that other examples and instances of violations
may have totally or partially different logic and rationale and hence
some of the following discussion and analysis should not apply to
them (i.e. this simple example does not typically represent all instances
and examples of claimed violation).}

Let us have a system of two charged particles (where we label them
as $q_{1}$ and $q_{2}$ for simplicity) which are initially at relative
rest. When one of these particles (say $q_{1}$) moves it should feel
the action force exerted by the field of $q_{2}$ immediately (because
it is located in this field) but $q_{2}$ does not feel the reaction
force exerted by the field of $q_{1}$ immediately (because the change
in its field propagates at a finite speed according to the mainstream
physics). So, while (at the instant of movement of $q_{1}$) there
is a force acting on $q_{1}$ (which is supposedly an ``action force''),
there is no force acting on $q_{2}$ (which is supposedly a ``reaction
force'') and this should represent a violation to Newton's third
law.

In fact, this supposed violation is based on the following implicit
assumptions (and possibly other assumptions noting that such assumptions
generally depend on the specific examples and instances as well as
the adopted theoretical frameworks):
\begin{enumerate}
\item The rejection of action at a distance (which is generally not acceptable
in the mainstream physics)\footnote{In fact, action at a distance gained some support in recent decades
(due mainly to seemingly supportive evidence from quantum physics
in general and quantum entanglement in particular) and hence it is
more accepted now.} and hence the interaction between the particles is taking place (supposedly)
through fields the propagation of signals through them is subject
to certain speed restrictions (which the mainstream physics impose
through the adoption of special relativity).
\item The existence of a privileged frame. This is because when we assume
$q_{1}$ moving (and $q_{2}$ at rest) it should be moving either
relative to an absolute frame (which is a privileged frame) or relative
to a frame in which $q_{2}$ is at rest (which is also a privileged
frame even if it is not an absolute frame). It is worth noting that
this assumption implies either the falsehood of special relativity
(due to the existence of absolute frame) or the dependence of the
violation of Newton's third law on the chosen frame since we can choose
a frame in which the two charged particles move symmetrically towards
each other and hence Newton's third law is not violated in this frame
(although it should be violated in other frames).
\item The action-reaction pair is actually between the two particles and
not between each particle and the field at its position. This is because
if the action-reaction pair is between each particle and the field
at its position then we have two separate action-reaction pairs where
in each pair the action of one agent (whether this agent is the particle
or the field at its position) is encountered by an instantaneous reaction
by the other agent with no delay or retardation. However, as we will
see this fix or proposal should not be accepted within the framework
of classical mechanics (because classical mechanics is a theory for
material particles not fields; moreover the paradigm of ``force acting
on a field'' seems bizarre in classical mechanics and inconsistent
with its conceptual framework even if we accept the paradigm of ``field''
in this mechanics).
\end{enumerate}
\vspace{0.1cm}The dependence of the claimed violations on certain
assumptions (as highlighted and exemplified in these points which
are related to this specific example) should highlight an important
and general issue that is, the validity of the claimed violations
generally depend on certain theoretical choices and preferences (most
of which are related to the commonly accepted doctrines in mainstream
modern physics). For example, if we accept action at a distance or
reject special relativity or accept the paradigm of ``force acting
on a field'' some of these claimed violations can vanish right away.

Anyway, according to the literature of mainstream physics (or at least
this is the common view in the literature), although Newton's third
law is (potentially) violated in situations like this, the principle
of conservation of momentum is not because the electromagnetic field
(or rather the ``ambient field'' to be more general) also carries
momentum (as will be discussed later on), and hence the total momentum
of the system (which consists of the two charges plus the electromagnetic
fields in the above example) is conserved despite the ``fact'' that
the total momentum of the system of the two charges alone (i.e. without
their fields) is not conserved due to the (potential) violation of
Newton's third law. However, alternative views can also be found
in the literature where (according to some of these views) neither
the conservation of momentum nor Newton's third law are violated (see
for instance \cite{Cullwick1952,Sebens2018}).

An important remark (which may be captured partially from the previous
discussion) is that these different (and possibly contradicting) views
are generally based on theoretical analysis and not on experimental
evidence (e.g. by direct measurement of action and reaction forces).
In fact, even those views which are supposedly based on experimental
evidence rely (mostly if not entirely) in their analysis and conclusions
on selected theoretical assumptions and hypothetical frameworks and
hence they are not actually (or at least purely) experimental. In
short, they depend in their validity and rationale on certain choices
and preferences ad hence they are not unconditionally experimental.

So, we can say (briefly and generally) that the alleged violations
(as well as the different and contradicting views about them) are
highly dependent on the adopted theoretical frameworks and hence in
most cases the disputes about these violations cannot be settled down
decisively and conclusively (e.g. by empirical evidence) noting as
well that no sufficient experimental efforts have been dedicated to
the investigation of these violations and disputes.

\section{\label{secInstancesViolation}Instances of Claimed Violation of Newton's
Third Law}

There are many claims in the literature of physics about violations
of Newton's third law in different physical systems and circumstances
and for various reasons. In the following subsections we investigate
common instances or cases of violation (or at least potential or tentative
or claimed violations) of Newton's third law noting that these violations,
or rather some of them, do not necessarily belong to different categories
although we categorize them in separate subsections for the sake of
clarity and organization.

\subsection{\label{subsecElectrodynamics}Violation in Electrodynamics}

According to the literature of mainstream physics, Newton's third
law is violated in electrodynamics (i.e. in some situations and circumstances).
For example, we read in Griffiths (see $\S$ 8.2.1 of \cite{GriffithsBook2013})
the following excerpt (noting that the \textit{italicization} is from
Griffiths):

In electro\textit{statics} and magneto\textit{statics} the third
law holds, but in electro\textit{dynamics} it does \textit{not}. Well,
that's an interesting curiosity, but then, how often does one actually
use the third law, in practice? \textit{Answer}: All the time! For
the proof of conservation of momentum rests on the cancellation of
internal forces, which follows from the third law. When you tamper
with the third law, you are placing conservation of momentum in jeopardy,
and there is hardly any principle in physics more sacred than that.
Momentum conservation is rescued, in electrodynamics, by the realization
that the \textit{fields themselves carry momentum}. This is not so
surprising when you consider that we have already attributed \textit{energy}
to the fields. Whatever momentum is lost to the particles is gained
by the fields. Only when the field momentum is added to the mechanical
momentum is momentum conservation restored. (End of quote)

So according to Griffiths (and actually many other scholars),\footnote{For example, we read in Purcell and Morin (see page 679 of \cite{PurcellM2013})
the following (noting that the \textit{italicization} is from the
authors): We see that Newton's third law, applied to the charges,
does \textit{not} hold. Equivalently (since $F=dp/dt$), the total
momentum of the proton plus pion is not conserved. However, the sacred
fact that is still true is that the total momentum of the \textit{entire
system} is conserved. And the system here consists of the two charges
\textit{plus} the electromagnetic field. We will learn in Chapter
9 that there is momentum \textit{in the field}, and the field is changing
here. The total momentum (proton plus pion plus field) is indeed conserved.
This is not a two-body system! (End of quote)\\
However, we should take notice of the condition ``applied to the
charges'' in this quote which may suggest that Newton's third law
still holds (like the conservation of momentum) to the entire system
(and this could be a difference with the view of Griffith). In fact,
the phrasing of Purcell and Morin may indicate their intention to
avoid taking a specific view about the violation of Newton's third
law (and this should reflect and highlight the problematic nature
of this issue).} we saved the law of conservation of momentum but lost Newton's third
law (noting that attributing part of the momentum to the field, which
supposedly saves the law of conservation of momentum, does not save
Newton's third law because the subject of this law is the forces on
the two particles with no reference to the field noting that classical
mechanics, and its Newtonian formulation in particular, is a mechanics
of particles not of fields). This simply means (based on the claim
and analysis of Griffiths):
\begin{enumerate}
\item Newton's third law is limited in validity and application, i.e. it
is \textit{sacred} but not as \textit{sacred} as the law of conservation
of momentum. This means a limitation on the validity of the formalism
(and hence the epistemology) of classical mechanics even within its
already-limited domain of validity (i.e. classical macroscopic scale
and inertial frames of reference; see \cite{SochiEpistClass12024}).
\item If the law of conservation of momentum is derived from Newton's third
law exclusively (at least in some fields and disciplines of physics)\footnote{\label{fnConMomExpTheo}It should be noted that although the principle
of conservation of momentum should acquire (generally and primarily)
its legitimacy and validity from experiment and observation, such
acquisition normally relies (at least partially) on theoretical considerations
and formulations (such as Newton's third law) as well and hence it
could be affected by such limitations (e.g. on the validity of Newton's
third law). In fact, this should apply to experimental and observational
evidence in general because no experiment or observation can be of
substantial use without a proper theoretical framework that is required
for its structuring and analysis.\\
We should also note that in the light of footnote \ref{fnConMomSym}
and the surrounding text, the derivation of the conservation of momentum
from Newton's third law should still be needed (at least in some disciplines
and circumstances) if such theoretical foundations are rejected or
questioned or restricted for some reason (e.g. due to a fundamental
position against such highly theoretical and mathematical methodologies
or because of certain technicalities which may be found in the literature).} then it is actually derived partly due to the partial validity of
Newton's third law (as seen in the previous point). This could mean
that the law of conservation of momentum (at least in classical mechanics)
is not as general as it should be because it is subject to the same
limitations of Newton's third law and hence it is limited to the instances
in which Newton's third law is valid. This could put some restrictions
on the validity and application of the law of conservation of momentum
(theoretically and practically) in classical mechanics and its applications
and extensions (and possibly even beyond classical mechanics). However,
we are not aware of such restrictions in the literature of mainstream
physics.
\item This violation of Newton's third law is not due to a restriction or
condition on the law by imposing certain conceptual or theoretical
restrictions on this law and its domain of validity and application
(because being in electrodynamics is not such a conceptual or theoretical
restriction) but because of the invalidity of this law in itself.
This means that this violation (if held) could invalidate Newton's
third law altogether and disqualify it as a law, and hence Newton's
third law is not really a law, i.e. it is no more than a useful rule
of thumb or a practical recipe for tackling and solving the problems
in classical mechanics and related fields of physics (or something
like these). This can have serious conceptual and theoretical consequences
not only on Newton's third law in classical mechanics but possibly
on the law of conservation of momentum as well (which is supposedly
derived from Newton's third law in classical mechanics and possibly
beyond; also see footnote \ref{fnConMomExpTheo}).
\item Referring to the previous point, we should draw the attention to a
misconception about this issue among some scholars who talk about
violation in classical mechanics and violation in electrodynamics
as if these violations are physically distinct and different, and
hence these scholars seem to allow violation to Newton's third law
in electrodynamics but not in classical mechanics (or at least they
put the onus and blame on electrodynamics specifically in these violations).
In this regard, we should say that the physical situations in electrodynamics
which are subject to the main limitations of classical mechanics (i.e.
classical macroscopic scale and inertial frames of reference; see
\cite{SochiEpistClass12024}) should be subject to the physics of
classical mechanics from a purely mechanical perspective (i.e. from
the perspective of the kinematic and dynamics of motion) even though
their particular physics (especially their dynamical aspects such
as forces and their origins) is based on another theory of physics
(i.e. electrodynamics in this case). In fact, this is exactly what
necessitated the use of other theories of physics to determine the
force in Newton's \textit{second} law which is considered (by some
scholars) as a definition for this reason (at least in part; see \cite{SochiEpistClass12024}).
In other words, the need for a physical theory (other than classical
mechanics such as electrodynamics) to determine the specific dynamical
agents and physical actors and how they operate within the given physical
situation does not exclude the given physical situation from the validity
domain of classical mechanics, and hence as long as the given physical
situation is within the domain of validity of classical mechanics
then it is a classical mechanical problem and thus it should be subject
(from mechanical perspectives and considerations kinematically and
dynamically) to the laws and principles of classical mechanics (i.e.
no violation of Newton's third law or any other law can be tolerated
as long as we accept classical mechanics as a valid physical theory
in that situation).\footnote{To support our argument about ``having a force law from electrodynamics
is not sufficient in itself to make the physical situation an electrodynamics
problem and not a classical mechanical problem'' we can argue that
in some examples of claimed violations in electrodynamics (and possibly
other disciplines) such as the example given in $\S$ \ref{subsecTypicalExample},
the rationale of violation applies equivalently (i.e. equivalent to
electrodynamics) to examples that belong exclusively to classical
mechanics (in its extended sense that includes gravity). For example,
if we replace the two charged particles in the example given in $\S$
\ref{subsecTypicalExample} with two gravitating massive particles
then the same logic of violation could apply in this case (which is
a problem that belongs to classical mechanics exclusively).} This is unlike the situation where the other theory used in the modeling
and formulation of the given physical situation is in contradiction
with the classical mechanics from the beginning and hence it cannot
be within the domain of validity of classical mechanics (as it is
the case with relativistic mechanics; see $\S$ \ref{subsecRelativistic});
in which case a violation of Newton's third law (or any other law
of classical mechanics) can be tolerated since the given situation
is not within the domain of validity of classical mechanics.
\end{enumerate}
\vspace{0.1cm}The attempts to save Newton's third law by certain
conceptual and technical manipulations (e.g. by Sebens \cite{Sebens2018}
who included the ``mass of field'' so that fields can be acted upon
by forces in, seemingly, a classical sense)\footnote{As indicated above, Sebens \cite{Sebens2018} ascribes mass to the
field in his attempt to save Newton's third law (and hence forces
can act on fields). However, this may be rejected (i.e. within this
context and purpose even if it is accepted in other contexts and purposes)
based on the fact that fields have no mass in a classical sense to
which classical mechanics (and Newtonian formulation in particular)
applies even if the concept of ``mass of field'' is accepted within
the framework of other theories of modern physics. In fact, the concept
of ``field'' (let alone the concept of ``mass of field'') does
not exist within the framework of Newtonian mechanics which is essentially
a mechanics of particles not fields. We should also refer the reader
to \cite{Cullwick1952} who seems to be the originator of this idea
(i.e. inclusion of ``mass of field'' to save Newton's third law).} do not seem consistent in spirit (if not in letter) with Newton's
third law (and classical mechanics in general). The least that can
be said in this regard is that (most if not all) these attempts are
not intuitive or make much sense within the framework of classical
mechanics and the Newtonian formulation in particular (and some may
not make much sense even beyond the framework of classical mechanics).
So, even if these attempts (or some of them) are accepted in general
(based for instance on their presumed validity and merit from a formal
perspective) they may save the principle of conservation of momentum
(but not Newton's third law and within classical mechanics), and this
could be limited to the formalism but not the epistemology (i.e. the
epistemology could be affected anyway).

Also, the attempts to save the conservation of momentum by similar
conceptual and technical manipulations are similarly problematic (although
the conservation of momentum is more fundamental and entrenched in
the fabric and structure of contemporary physics than Newton's third
law and has more theoretical and experimental support independent
of Newton's third law and hence it may be saved mainly for this reason).
As indicated earlier, the inclusion of field (to save the principle
of conservation of momentum) my be rejected (at least within the framework
of classical mechanics and the Newtonian formulation in particular)
because the Newtonian formulation of classical mechanics is a theory
for material particles and not for fields and hence the inclusion
of field (even if it is accepted in principle and within other theories
and branches of physics) may save the conservation of momentum outside
classical mechanics but not inside classical mechanics (and its Newtonian
formulation in particular).

So in brief, if we accept the rationale of the claimed violation of
Newton's third law in electrodynamics\footnote{In fact, the following argument may apply to similar instances of
violation of Newton's third law which will be investigated in the
next subsections, i.e. it is not restricted to electrodynamics.} (as outlined earlier) then we may lose the conservation of momentum
(at least from this basis and foundation), as well as Newton's third
law, within classical mechanics or at least within its Newtonian formulation
(noting that the other formulations of classical mechanics do not
include Newton's third law explicitly and directly although it may
be obtained by derivation from their principles or by their presumed
equivalence to the Newtonian formulation as indicated earlier; see
\cite{SochiEpistClass12024}). Regarding outside classical mechanics
(assuming that the validity of Newton's third law extends beyond classical
mechanics, as some physicists believe), it seems that we will lose
Newton's third law in its intuitive and ``classical'' sense at least
(and hence we may lose it altogether as declared explicitly by several
scholars some of whom are cited in this paper). We may also lose the
conservation of momentum although this should depend on a number of
factors (such as the embraced theoretical framework which usually
depends on the branch of physics and the personal choices and preferences)
and is usually excluded due to the fundamental role of the conservation
of momentum in contemporary physics and its extra sources of support
and validation (both theoretical and experimental) as indicated earlier.

Anyway, any loss (whether to Newton's third law or to the conservation
of momentum and whether within or outside classical mechanics) should
put big question marks on Newton's third law epistemologically (and
possibly formally) as a law in itself and on its relation to the conservation
of momentum. Some of these question marks may also extend to the conservation
of momentum (although this does not seem very likely noting that the
conservation of momentum is more central and essential than Newton's
third law in contemporary physics as reflected by the common view
among the mainstream physicists which accepts the violation of Newton's
third law but not the violation of the conservation of momentum).

\subsection{\label{subsecRelativistic}Violation in Relativistic Mechanics}

According to the literature of mainstream physics, Newton's third
law does not hold in relativistic mechanics (i.e. in some situations
and circumstances). For example, we read in Taylor (see page 21 of
\cite{TaylorBook2005}) the following excerpt (noting that the \textit{italicization}
is from Taylor):

Within the domain of classical physics, the third law, like the second,
is valid with such accuracy that it can be taken to be exact. As speeds
approach the speed of light, it is easy to see that the third law
cannot hold: The point is that the law asserts that the action and
reaction forces, $\mathbf{F}_{12}(t)$ and $\mathbf{F}_{21}(t)$,
\textit{measured at the same time t}, are equal and opposite. As you
certainly know, once relativity becomes important the concept of a
single universal time has to be abandoned --- two events that are
seen as simultaneous by one observer are, in general, \textit{not}
simultaneous as seen by a second observer. Thus, even if the equality
$\mathbf{F}_{12}(t)=-\mathbf{F}_{21}(t)$ (with both times the same)
were true for one observer, it would generally be false for another.
Therefore, the third law cannot be valid once relativity becomes important.
(End of quote)

Similarly, we read in Griffiths (see $\S$ 12.2.4 of \cite{GriffithsBook2013})
the following excerpt (noting that the \textit{italicization} is from
Griffiths):

Unlike the first two, Newton's \textit{third} law does not, in general,
extend to the relativistic domain. Indeed, if the two objects in question
are separated in space, the third law is incompatible with the relativity
of simultaneity. For suppose the force of $A$ on $B$ at some instant
$t$ is $\mathbf{F}(t)$, and the force of $B$ on $A$ at the same
instant is $-\mathbf{F}(t)$; then the third law applies \textit{in
this reference frame}. But a moving observer will report that these
equal and opposite forces occurred at \textit{different times}; in
his system, therefore, the third law is violated. Only in the case
of contact interactions, where the two forces are applied at the same
physical point (and in the trivial case where the forces are constant)
can the third law be retained. (End of quote)

Similar views and stands can be found in the literature of mainstream
physics. In fact some of these views sound even stronger than these
(for example, according to French \cite{FrenchBook1968}: ``one of
Newton's basic assertions about forces between bodies - the equality
of action and reaction - has almost no place in relativistic mechanics'').
Moreover, these views generally vary in the cause of violation of
Newton's third law in relativistic mechanics, e.g. whether it is because
of the relativity of simultaneity or because of the denial of action
at a distance and the speed limit in communication (noting that these
causes are generally interrelated) although this should depend in
part on the specific example of violation as well as on the personal
choices and opinions which the analysis depends on.

Anyway, the violations of Newton's third law in relativistic mechanics
may be challenged by various arguments which are normally based on
different theoretical grounds and they usually depend on one's own
convictions. For example, the challenges may be based on the rejection
of special relativity or some of its postulates or principles or implications
(such as the denial of action at a distance and the speed limit in
communication), or based on the introduction of certain modifications
and interpretations on the relativistic mechanics which lead to the
negation of such violations.

However, all these claimed violations and their challenges are based
on their own theoretical frameworks which depend primarily on personal
choices and preferences, and hence it is virtually useless to try
to prove or disprove any one of these claims or their challenges in
an absolute sense and meaningful manner (refer to $\S$ 2.2 of \cite{SochiKetab1}
for more details). Nonetheless, we can say (in a more general and
useful argument away from the specific and detailed technicalities
which are required to deal with these violations and challenges specifically
and individually) that Newton's laws are supposedly limited by the
``scale'' factor (related to speed) to the classical domain (see
\cite{SochiEpistClass12024}) and hence any supposed failure of Newton's
third law in relativistic mechanics should not be a problem (at least
in itself) because the domain of validity of relativistic mechanics
is not included in the domain of validity of classical mechanics.\footnote{However, we should remark that (at least) some examples of claimed
violations in relativistic mechanics are based on faulty arguments
or questionable rationale, and hence they should be rejected regardless
of the acceptability (in principle) of such violations due to the
imposed limitation on the domain of validity of classical mechanics
and Newton's laws of motion.} Yes, such supposed violations should be addressed by those physicists
who believe that the validity of Newton's third law extends beyond
classical mechanics (assuming that these physicists believe in relativistic
mechanics).

Anyway, if this violation of Newton's third law implies a violation
of the law of conservation of momentum (which Newton's third law supposedly
implies even in relativistic mechanics, assuming the validity of Newton's
third law beyond classical mechanics and hence it extends in principle
to relativistic mechanics) then we could have a more serious problem
because momentum is supposedly conserved in relativistic systems (although
this should require some modification to the definition of momentum
and possibly other related modifications in conceptualization and
formulation). However, these issues and details belong to the relativistic
mechanics (which is not the subject of investigation of the present
paper) but we should remember what we said earlier that is the conservation
of momentum in contemporary physics has extra sources of support and
validation (both theoretical and experimental) which are independent
of Newton's third law and hence the conservation of momentum should
be saved eventually even if Newton's third law is violated or lost
altogether.

\subsection{\label{subsecViolationNonInertial}Violation in Non-Inertial Frames}

Newton's third law is violated (according to some scholars) by the
inertial (or fictitious or pseudo) forces which appear in non-inertial
frames. We may quote in this regard the following passage (see page
60 of \cite{Irodov2002}):

Inertial forces are caused not by the interaction of bodies, but by
the properties of non-inertial reference frames themselves. Therefore
inertial forces do not obey Newton's third law. (End of quote)

We discuss and assess this issue in the following remarks:
\begin{enumerate}
\item If these forces are really fictitious (in the sense of being imaginary
and illusory) then there should be no violation of any law because
they are fictitious and hence it is nonsensical to attribute any real
physical effect (such as violating Newton's third law) to them. So,
anyone who accepts this claim of violation should accept in advance
the physical reality of fictitious forces; otherwise he should be
contradicting himself (at least implicitly).
\item Newton's laws of motion are restricted in validity to inertial frames
of reference. Now, fictitious forces (whether we consider them real
or not) appear only in non-inertial frames of reference and hence
any supposed violation of Newton's third law by these forces should
not be a problem to Newton's third law or to the Newtonian mechanics
in general because these forces are not supposed to be subject to
the Newtonian laws or mechanics because they are not within its domain
of validity. Yes, if we consider Newton's third law (or rather its
essence) as a fundamental law of Nature that should (in principle)
apply in all frames of reference (as some physicists believe) then
this could be a sensible claim of violation (regardless of its merit
and correctness from this aspect or other aspects and regardless of
its relation to classical mechanics).
\item It may be argued that if these fictitious (or rather inertial) forces
are real then what is the problem in having an interaction (thanks
to the inertia) between the material objects and the \textit{physical}
space where one of them acts while the other reacts (even though this
may not be consistent with the Newtonian view and possibly other views).
In fact, we may even conceptualize this interaction and label it specifically
as ``inertial interaction''.\\
However, this can be challenged by the fact that we indicated earlier,
i.e. Newtonian mechanics is essentially a theory for the mechanics
of particles and hence any supposed interaction that is to be subject
to Newton's third law must be between massive particles, and the space
is obviously not a particle (in fact this seems to be indicated in
the above quoted passage of \cite{Irodov2002} by ``interaction of
\textit{bodies}'').\footnote{In fact, this challenge may be enforced by claiming that no ``tangible
effect'' on space can be observed in such presumed interaction and
hence this presumed interaction cannot be real.}\\
We should note that this reason may also be used as an excuse or pretext
for justifying the ``non-violation'' by claiming that since the
supposed interaction involves a non-particle entity (i.e. space) then
it should be outside the domain of validity (or rather outside the
formulation) of Newtonian mechanics and Newton's third law in particular\footnote{In fact, this may also be used as a reason for ``violation'' because
it is outside the ``domain of validity'' of Newtonian mechanics.} (but we should notice as well that this sort of argument may be applied
even to the real forces in the supposed interactions between particles
and physical fields, such as the electromagnetic field, which we discussed
earlier).
\end{enumerate}

\subsection{Violation in Interaction between Space and Material Objects}

This alleged violation of Newton's third law is based on a philosophical
interpretation to the ``Newtonian epistemology'' about the nature
of the relationship between the Newtonian space and the material objects
that it ``contains'' (which, i.e. the relationship, supposedly provides
an explanation to the inertial effects that originate from the Newtonian
space). This alleged violation is supposedly expressed in the following
quote from the Foreword of \cite{JammerBook1993} which reads:

The concept of space was enriched and complicated by Galileo and Newton,
in that space must be introduced as the independent cause of the inertial
behaviour of bodies if one wishes to give the classical principle
of inertia (and therewith the classical law of motion) an exact meaning.
To have realized this fully and clearly is in my opinion one of Newton's
greatest achievements. In contrast to Leibniz and Huygens, it was
clear to Newton that the space concept (a) was not sufficient to serve
as the foundation for the inertia principle and the law of motion.
He came to this decision even though he actively shared the uneasiness
which was the cause of the opposition of the other two: space is not
only introduced as an independent thing apart from material objects,
but is also assigned an absolute role in the whole causal structure
of the theory. This role is absolute in the sense that space (as an
inertial system) acts on all material objects, while these do not
in turn exert any reaction on space. (End of quote)

We may also quote in this regard the following excerpt  from \cite{AssisG1995}:
In Newton's mind this change of condition was acceleration of the
body relative to absolute space. It is puzzling that Newton never
mentioned the fact that the force of inertia, which opposes acceleration,
violates his Third Law of Motion, for absolute space cannot sustain
the required reaction force. (End of quote)

In fact, this may be seen as a bizarre view and twisted interpretation
to the essence of the relationship between space and material objects
in classical mechanics if it is supposed to be about Newton's third
law (as the second quote states explicitly). However, regardless of
this the essence of this view does not seem to be connected to Newton's
third law in the technical sense of this law because this is a view
taken from a purely philosophical perspective (i.e. it is not physical
or technical) and hence this alleged violation (regardless of its
soundness) is irrelevant to our subject and objective of the present
paper. Moreover, this sort of interaction may be considered as being
cosmological in scale and hence it is outside the domain of validity
of classical mechanics and its Newtonian formulation (see \cite{SochiEpistClass12024})
because the Newtonian space is a cosmological entity and not a physical
entity in the classical macroscopic sense and at this level.

We should also note (mainly in connection to the second quote) that
if the inertial forces are considered as reaction forces generated
by the Newtonian \textit{physical} space in response to the action
forces that generate the acceleration (see \cite{SochiEpistClass12024})
then it may be claimed that there should be no violation to Newton's
third law. However, this may be challenged by the fact that both forces
(according to this conceptualization) act on the same object and hence
this is not a subject for Newton's third law whose (action and reaction)
forces act on separate objects. So, if ``no violation'' should be
accepted then it should be in this negative sense.

We should finally note that the issue in this subsection is intimately
related to the issue of the previous subsection (see $\S$ \ref{subsecViolationNonInertial}).
However, we prefer to consider these two issues as separate and independent
of each other. This is based on our distinction between the ``resistance
of inertia'' (or the ``force of inertia'' as stated in the previous
quote) which is a property of material objects that demonstrates itself
in all types of frame (whether inertial or non-inertial), and the
``inertial forces'' (or fictitious or pseudo forces) which are a
property of non-inertial frames of reference since these forces demonstrate
themselves only in non-inertial frames.

\subsection{\label{subsecViolationBiology}Violation in Biology}

There are claims in the recent literature of physics (mostly within
the category of science popularization) that Newton's third law is
violated in some biological systems and phenomena such as in the movement
of sperms. However, we think these claims should not be taken seriously
at this stage for various reasons such as:
\begin{enumerate}
\item There are many ambiguities surrounding these claims and this should
question their credibility. It is likely that these claims are made
by biologists who have poor understanding of physics and its main
role which is the rigorous determination of the fundamental laws of
Nature and hence it is not about phenomenological analogies and similarities
that can be found in popular science and everyday chats and activities.
In fact, this claim of violation may be similar to the use of Newton's
third law in daily conversations and political debates (where some
people try to give themselves the image of being smart, sophisticated
and well-educated by quoting Newton's third law foolishly and nonsensically).
\item This issue is related to the issue of biology (and life ultimately)
and if it is part of physics or not, i.e. whether biological phenomena
and life obey the physical laws (possibly in a very complex form)
and hence they belong to physics or not (in which case they should
be subject to a different type of laws and hence they are beyond the
reach of physics at least in its current state of development). This
issue, as well as its consequences and implications, is not clear
cut and could be a matter of choice and convention. Therefore, this
issue may not have a definite answer since it depends on personal
choice and preference.\\
However, it should be obvious that everything that we observe in this
world belongs to the physical world including the biological systems
and phenomena (to which these cases of alleged violation of Newton's
third law belong) and hence they are physical in this sense and should
be subject ultimately to the fundamental laws of physics. Nevertheless,
it should also be obvious that not everything that is physical (i.e.
belonging to the physical world and subject ultimately to the fundamental
laws of physics) belongs to the science of physics. In fact, contemporary
physics should be restricted by definition and convention to phenomena
that do not involve things like life and human behavior which are
too complex to be tackled by our current (and rather limited) knowledge
and understanding of the physical world. Hence, biological and social
sciences for instance (as well as many other branches of science and
knowledge) should not be seen as branches of physics (at least in
its current state) despite being physical in the aforementioned sense
and subject ultimately to the fundamental laws of physical world.\\
Yes, if physics becomes so powerful and broad to the limit that it
can formulate laws and principles capable of describing and predicting
the behavior of biological and sociological systems and their phenomena
(as well as similar complex systems and phenomena) then these branches
of science could be included as branches of physics in its extended
sense. However, this seems to be a non-achievable (and possibly impossible)
goal at least for the foreseeable future. If our current physics (which
is based on many simplifications and deals mostly with very simple
physical systems) cannot find (exact) solutions even to some of the
simplest systems (like the helium atom) then there is no hope (at
least for the foreseeable future) to extend physics to this ambitious
extent and dimension. In fact, physics in its current state (despite
its complexity and elaboration) is a very simple and primitive science
(in comparison to the complexities of the physical world that it is
supposed to depict) that effectively deals only with the very simple
systems using very simple models, patterns, techniques, methodologies
... etc. most of which are no more than imitations and approximations.
So, we should not be very ambitious at this stage of scientific development
to expect that we can develop physical theories about life or biological
systems or society or economy for instance.\\
To conclude, the investigation and analysis of biological systems
and phenomena by contemporary physics and within its current frameworks
and tools is entirely inappropriate. Contemporary physics is too simple
to deal with such complex systems and phenomena even though these
systems and phenomena are physical in the above sense and they belong
to the physical world and hence they should be subject ultimately
to the fundamental laws of ``physics'' although this must be in
a very complex and sophisticated way due to their complexity and sophistication.
So, any claim of violation of Newton's third law (or any other physical
law) in biological systems and phenomena should be treated at this
stage with caution and skepticism.
\end{enumerate}

\subsection{Other Violations}

There are a number of other claims (mostly in the recent literature
of physics) of violations of Newton's third law in various physical
systems and disciplines. In fact, some of these systems and disciplines
do not seem to belong to classical mechanics and may not fall within
its domain of validity. Moreover, some of these claimed violations
may not be really violations to the fundamental physics of Newton's
third law (i.e. they can be similar in nature to the claimed violations
in biology which we discussed in $\S$ \ref{subsecViolationBiology}).
These claims include, for instance, violations in colloidal systems,
fluid dynamical systems, and statistical mechanical systems (see for
example \cite{DzubiellaLL2003,HayashiS2006,BuenzliS2008,IvlevBHDNL2015}
noting that there is some overlapping in this categorization).

However, such violations may be explained and justified by the commonly
held view of keeping the (very sacred) principle of conservation of
momentum while sacrificing the (less sacred) Newton's third law (if
this sacrification is required). They may also be explained and justified
by claiming, for instance, that Newton's third law holds true for
the complete ``particles-plus-environment'' systems (similar to
the previous justifications with regard to the conservation of momentum)
even though it does not hold for the particles (or matter) involved
in these violations (see for example \cite{IvlevBHDNL2015}). In fact,
claims like this are generally based on a desire to keep Newton's
third law safe, i.e. they are not based on solid experimental evidence
or theoretical foundation independent of this desire (which is generally
based on a conviction that Newton's third law is a fundamental law
due, mostly, to its intimate link to the conservation of momentum
which is seen unanimously as a fundamental law of Nature).

\section{General Remarks}

Before we reach the end of this paper by summarizing our conclusions
it is useful to take notice of some general remarks as outlined in
the following subsections.

\subsection{\label{subsecDependOnFramework}Dependence of Claimed Violations
on Adopted Theoretical Frameworks}

The views and stands on both sides of the debate about the claimed
violations of Newton's third law (as well as their potential impacts
and implications related, for instance, to their effects on the integrity
and generality of the principle of conservation of momentum) are highly
dependent on the adopted theoretical frameworks and individual choices
as well as disputable theories and paradigms (such as special relativity
and action at a distance). In fact, in most cases the judgments made
and the conclusions reached about these alleged violations and their
potential consequences strongly depend on individual views and opinions
(which mostly represent personal tastes and preferences) about certain
physical theories and issues (e.g. whether we believe in special relativity
or not and whether we employ Lorentz force law in our theoretical
framework and analysis or not).

We should also note that certain conceptualization elements which
are closely related to epistemology (rather than formalism) are employed
sometimes in analyzing the instances of claimed violation and drawing
the conclusions from them. For example, in some instances of alleged
violation of Newton's third law the authenticity of violation depends
(in part) on the conceptualization of the reaction force which (i.e.
this conceptualization) depends on the missing asymmetry in Newton's
third law (which we discussed in $\S$ 2.3.6 of \cite{SochiEpistClass12024}),
i.e. whether we have a symmetric situation (since the force exerted
by both objects can be conceptualized as only action or only reaction)
or we have an asymmetric situation\footnote{Or rather a double or reciprocal asymmetric situation.}
(since we can conceptualize the situation as having two action forces
and two reaction forces which is like the situation of two persons
pushing each other equally and symmetrically); see points 2 and 3
of footnote 20 of \cite{SochiEpistClass12024}.

\subsection{\label{subsecInsufficiencyExperiment}Insufficiency of Experimental
Evidence in this Debate}

\noindent Most views and opinions about the claimed violations of
Newton's third law (on both sides of the debate) are not based on
hard experimental evidence, i.e. they are based either on purely theoretical
analysis (which usually rests on commonly accepted theories and paradigms
of modern physics) or on experimental evidence analyzed by tentative
or questionable theoretical frameworks (see $\S$ \ref{subsecDependOnFramework}).
This should cast a shadow on the credibility of many of these instances
of violation and the adopted views and stands (on both sides of this
debate).

In fact, some of these claims can be verified conclusively and decisively
by conducting laboratory experiments to measure the (magnitude, direction
and alignment of) action and reaction forces directly and independently
of the adopted theoretical frameworks or theories (such as electrodynamics
and special relativity) which are used in the analysis of the alleged
violations and in the drawing of the reported results and conclusions.

\subsection{Seriousness of Claimed Violations}

In our view, the claimed violations (assuming their authenticity,
i.e. the evidence and arguments forwarded in their support are correct
and acceptable in general) fall into three main categories:
\begin{enumerate}
\item Very serious violations which are (mostly) those that threaten the
principle of conservation of momentum (whether linear or angular)
or its generality from some aspects or in some fields. We should also
include in this category those which are threatening some fundamental
theories (such as special relativity) for those scholars who embrace
these theories.
\item Serious violations which are those that threaten Newton's third law
but not the principle of conservation of momentum (or other fundamental
principle or law or theory of physics).
\item Minor or trivial violations which are those that can be tackled and
fixed by some minor modifications and mild compromises and amendments
to the adopted frameworks and personal choices and preferences, and
hence any supposed breach or violation of Newton's third law (or other
principles or laws or theories of physics) can be avoided.
\end{enumerate}
\vspace{0.1cm}In fact, all these categories seem to exist within
the existing instances of claimed violations. It should be obvious
that the (degree of) seriousness of these claimed violations and breaches
should (similarly and equally) impact the epistemology of the affected
laws and theories (and actually the contemporary physics in general)
as well as the affected formalism.

\section{Conclusions}

We outline in the following points the main conclusions that we can
obtain from the investigation of the present paper:

\noindent $\bullet$ The violation of Newton's third law is mostly
(but not unanimously) accepted in contemporary physics. In fact, we
can identify three main groups of physicists with regard to this issue:
those who definitely accept this violation, those who definitely reject
this violation (and hence they propose certain remedies and fixes
to keep the third law intact), and those who do not take any specific
position and hence they avoid expressing their view, if they have
any view, about this issue (i.e. they do not seem to care about Newton's
third law as long as the conservation of momentum is preserved). This
is unlike the principle of conservation of linear and angular momentum
which (unanimously) survives in these supposed violations (i.e. we
are not aware of physicists who regard these supposed violations of
Newton's third law as violations to the conservation of momentum).

\noindent $\bullet$ The violation of Newton's third law should cast
a shadow on the status of this law as a real physical law within and
outside classical mechanics (noting that what is significant, or more
significant, of such violations is the violation within classical
mechanics due to the limited validity of Newton's laws to the validity
domain of classical mechanics, i.e. classical macroscopic scale and
inertial frames of reference).\footnote{\noindent We should note that ``limited validity of Newton's laws''
should mean ``in principle and primarily'' noting that (at least)
some of these laws may be extended in validity (according to the view
of some physicists) beyond classical mechanics and its domain of validity
(as explained earlier).} This should have serious epistemological (and possibly formal) implications
and consequences on classical physics (and possibly beyond).

\noindent $\bullet$ The violation of Newton's third law requires
further investigation to its actual relationship with the principle
of conservation of momentum and if this principle could be affected
or violated by the violation of Newton's third law (at least within
the framework of classical mechanics). This is because the aforementioned
unanimity about the integrity of the principle of conservation of
momentum in these instances of violation seems to be based (at least
in part) on a general conviction among contemporary physicists of
the sacredness of this principle which mostly originates from general
theoretical and mathematical arguments (such as the supposed properties
of space and related symmetries) rather than independent and specific
analysis and investigation to these instances of violation individually
and specifically.

\noindent $\bullet$ The violation of Newton's third law may require
imposing further limitations on the domain of validity of classical
mechanics (and its Newtonian formulation in particular) in addition
to those already-imposed limitations (i.e. classical macroscopic scale
and inertial frames of reference; see \cite{SochiEpistClass12024}).

\noindent $\bullet$ The contradicting views about the violation of
Newton's third law add more ambiguities (in addition to what we discussed
in \cite{SochiEpistClass12024}) to this alleged law especially from
the epistemological and interpretative perspectives.

\noindent $\bullet$ Most (if not all) views about the violation of
Newton's third law (on both sides of the debate) are not based on
hard experimental evidence, i.e. they are based either on purely theoretical
analysis or on experimental evidence analyzed by tentative or questionable
or debatable theoretical frameworks. Moreover, the judgments about
such violations usually depend on views and opinions (as well as personal
choices and preferences) about certain physical theories and issues
(e.g. whether or not we believe in special relativity or employ Lorentz
electrodynamics in our framework and analysis). These factors and
complications (as well as the other mentioned factors in this regard)
should increase the ambiguity and uncertainty about this problematic
issue.

\pagebreak{}

\phantomsection 
\addcontentsline{toc}{section}{References}\bibliographystyle{unsrt}
\bibliography{Bibl}

\end{document}